\documentclass[aps,prl,floatfix,twocolumn,superscriptaddress]{revtex4-1}
\usepackage{graphicx}
\usepackage{amsmath}
\usepackage{amssymb}
\usepackage{xr-hyper}
\usepackage{braket}

\usepackage{array,color}
\usepackage{multirow}

\usepackage{xcolor}

\newcommand{\NIST}{
Time and Frequency Division, National Institute of Standards and Technology, Boulder, CO, USA}
\newcommand{\CU}{
Department of Physics, University of Colorado, Boulder, CO, USA}
\newcommand{\UD}{
Department of Physics and Astronomy, University of Delaware, DE, USA}
\newcommand{\JQI}{
Joint Quantum Institute, NIST and the University of Maryland, College Park, MD, USA}
\newcommand{\UCLA}{
Department of Physics and Astronomy, University of California, Los Angeles, CA, USA}
\newcommand{\NUL}{
Department of Physics and Electronics, National University of Lesotho, Roma, LS}

\newcommand{\ThIV}{$^{229}$Th$^{3+}$\:}
\newcommand{\ave}[1]{\left< #1 \right>}

\begin{document}
\setlength{\parskip}{0pt}

\title{Quantum metrology algorithms for dark matter searches with clocks}

\author{M. H. Zaheer}
\email{hani@udel.edu}
\affiliation{\UD}
\author{N. J. Matjelo  }
\affiliation{\NUL}
\author{D. B. Hume}
\affiliation{\NIST}
\author{M. S. Safronova}
\affiliation{\UD}
\affiliation{\JQI}
\author{D. R. Leibrandt}
\email{leibrandt@physics.ucla.edu}
\affiliation{\UCLA}
\affiliation{\NIST}
\affiliation{\CU}

\date{\today}

\begin{abstract}

Quantum algorithms such as dynamical decoupling can be used to improve the sensitivity of a quantum sensor to a signal while suppressing sensitivity to noise.  Atomic clocks are among the most sensitive quantum sensors, with recent improvements in clock technology allowing for unprecedented precision and accuracy. These clocks are highly sensitive to variations in fundamental constants, making them ideal probes for local ultralight scalar dark matter. Further improvements to the sensitivity is expected in proposed nuclear clocks based on the thorium 229m isomer. We investigate the use of various quantum metrology algorithms in the search for dark matter using quantum clocks. We propose a new broadband dynamical decoupling algorithm and compare it with quantum metrology protocols that have been previously proposed and demonstrated, namely differential spectroscopy and narrowband dynamical decoupling. We conduct numerical simulations of scalar dark matter searches with realistic noise sources and accounting for dark matter decoherence. Finally, we discuss an alternative thorium nuclear transition excitation method that bypasses the technical challenges associated with vacuum ultraviolet lasers.

\end{abstract}

\maketitle

\section{Introduction}

The extraordinary improvement in the precision of atomic clocks over the last 15 years \cite{LudBoyYe15,BreCheHan19} has put clocks in an uncharted territory in which the fundamental postulates of modern physics are untested. For example, if fundamental constants, such as the fine structure constant $\alpha$, are space-time dependent, then so are the atomic, molecular, and nuclear transition frequencies. Variations of fundamental constants would change the ticking rate of clocks and make them depend on the location, time, or type of clock, since the frequencies of different clocks depend differently on the fundamental constants.

Ultralight scalar fields arise in many theories beyond the Standard Model \cite{BanGilSaf22,antypas2022new} and may source variations of fundamental constants that affect atomic clock frequencies \cite{DerPos14,Arv15}. Such particles can also be dark matter (DM), which makes 85\% of all matter in the Universe, but it is of yet unknown origin. In our galaxy, such dark matter exhibits coherence and behaves like a wave with an amplitude $\sim \sqrt{\rho_{\textrm{DM}}}/m_{\textrm{DM}}$, where $\rho_{\rm{DM}}=0.4 \: \mathrm{GeV/cm}^3$
is the local DM density and $m_{\textrm{DM}}$ is the DM particle mass \cite{Arv15}. The coupling of such DM to the Standard Model leads to oscillations of fundamental constants and, therefore, clock transition frequencies.
Since different clocks have different sensitivities to such effects, measurements of the ratio of two clock frequencies over time can be used to extract an oscillation signal at the DM Compton frequency for a wide range of DM masses and interaction strengths \cite{Arv15,Til15,Hee16}.
Transient changes in fundamental constants that are potentially detectable with networks of clocks may be induced by dark matter objects with large spatial extent, such as stable topological defects \cite{DerPos14,StaFla2015,WsiMorBob16,RobBleDai17,KalYu17}.
These advances started a new era of clock experiments aimed at detecting dark matter, with several ongoing searches \cite{antypas2022new}.

It was recently suggested that ultralight DM can form gravitationally bound objects that may be trapped by an external gravitational potential, such as that of the Earth or the Sun \cite{Banerjee:2019epw}.
It was shown in \cite{YuDai22} that such a halo can have extremely large DM overdensities, up to 17 orders of magnitude at 0.1 AU from the Sun, drastically increasing discovery potential. \citet{YuDai22} proposed a clock-comparison satellite mission with
two clocks onboard, to the inner reaches of the solar system to search for a dark matter halo bound to the Sun and look for the
spatial variation of the fundamental constants associated with a change in the gravitational potential. However, in this scenerio, the DM distribution peaks for about 10$^{-13}$ eV DM masses, at which clocks lose sensitivity due to the probe duration of the clocks. In this work, we explore different quantum algorithms for clock operation and propose a broadband dynamic decoupling scheme that allows one to reach such masses without significant losses of sensitivity.

The sensitivity of an optical clock to any variation of $\alpha$ (temporal, spatial, slow drift, oscillatory, gravity-potential dependent, transient, or other) is quantified
by the dimensionless enhancement factor $K$. Clocks with the best stability, total systematic uncertainty, and the highest possible
values of $\Delta K=K_1-K_2$ for clocks 1 and 2 have the highest discovery potential.
The  $K$ factors are small, $0.008-1.0$, for all current clocks based on optical lattices and most trapped ions clocks, with the exception of the Hg$^+$ quadrupole ($K=-3$) and Yb$^+$ octupole clock ($K=-6$) transitions \cite{DzuFla09}.
Clocks based on nuclear transitions can have much higher sensitivies to variations in the fundamental constants, due to considerably higher energies of transitions.

The transition frequencies of nuclear energy levels are well outside the laser-accessible range, by 4-6 orders of magnitude, with a single exception of $^{229\rm{m}}$Th.
Designing a clock based on this ultra-narrow nuclear transition \cite{PeiTam03,CamRadKuz12}, with a wavelength measured to be $148.81(41)$~nm (weighted mean of the measurements reported in Refs.~\cite{seiferle2019energy,Sikorsky2020energy,Kraemer2022energy}), is particularly attractive due to its insensitivity to many systematic effects. This transition is predicted to have a very large sensitivity to variations of the fundamental constants $\alpha$ and $m_q/\Lambda_{\textrm{QCD}}$ \cite{Fla06a,BerFla11}, where $m_q$ are the quark masses and $\Lambda_{\textrm{QCD}}$ parameterizes the strength of the strong force, several orders of magnitude larger than all present atomic clocks. The $K$ factor for the proposed Thorium nuclear clock is of the order of $10^4$ \cite{fadeev2020sensitivity}.

Here, we consider the linear coupling of ultralight scalar dark matter $\phi$ with the photon field given by the interaction Lagrangian \cite{Arv15}
\begin{equation}
\mathcal{L_{\phi\gamma}} = \kappa\phi \frac{d_e}{4e^2} F^{\mu\nu}F_{\mu\nu} \ ,
\end{equation}
\noindent where $\kappa=\sqrt{4\pi}/M_{Pl}$ is a dimensionless coefficient, $e$ is the electron charge, and $d_e$ is the dark matter QED coupling constant.

Typically, optical clocks are operated in a manner that is optimized to reach the highest possible precision for measuring static frequency differences or ratios at asymptotically long averaging times.  This is achieved by probing the clock transition using Rabi or Ramsey spectroscopy with as long of a probe duration as possible while maintaining coherence between the clock laser and the atoms.  The fundamental limit to the measurement precision is set by quantum projection noise and for clocks based on unentangled atoms this limit is referred to as the standard quantum limit (SQL).  For several highly-forbidden clock transitions, the atomic coherence time can be orders of magnitude longer than that of the best present-day clock lasers, so laser noise sets the experimental limit to measurement precision.

For frequency ratio measurements between an optical lattice clock and a trapped ion clock, a new technique called differential spectroscopy (DS) has recently been proposed \cite{Hume2016} and demonstrated \cite{Kim2022} which circumvents the laser coherence time limit to the probe duration.  In differential spectroscopy, the two clocks are probed synchronously using lasers that are phase locked to a common frequency comb, such that the correlated laser noise can be circumvented.  The probe duration of the ion clock can be extended all the way to its atomic coherence time if the lattice clock is operated with two atomic ensembles probed antisynchronously to achieve a zero-dead-time measurement of the laser phase (see Fig.~\ref{fig:broadbandDD}a).  Zero-dead-time differential spectroscopy is a promising technique for ultralight scalar dark matter searches at the low particle mass limit which would manifest as slow oscillations in the frequency ratio between thorium nuclear transition or ytterbium octupole transition trapped-ion clocks with a large sensitivity to the DM field, and strontium or ytterbium optical lattice clocks with a small sensitivity that would serve as the stable reference.  As shown in the right panel of Fig.~\ref{fig:broadbandDD}a, clocks operated using differential spectroscopy (and also standard Rabi or Ramsey spectroscopy) are primarily sensitive to oscillations at frequencies lower than the reciprocal of the clock probe duration, with the sensitivity decreasing rapidly with increasing frequency above this cutoff.  While the sensitivity at frequencies higher than the inverse probe duration can be improved by reducing the probe duration, this comes at the cost of higher QPN.

The fundamental challenge in quantum sensing is to increase the sensitivity to a signal while reducing the susceptibility to both technical noise such as laser noise and fundamental noise such as QPN.
Dynamical decoupling is one method to approach this problem. Ramsey pulse sequences with $\pi$ pulses inserted into the free evolution time (see Fig.~\ref{fig:broadbandDD}b) are used to dynamically decouple noise from the signal in a quantum system.
The first use of dynamical decoupling was in nuclear magnetic resonance, with the spin echo effect \cite{hahn1950spin}.
It was introduced in quantum information research as a means to deal with decoherence and dissipation \cite{viola1998dynamical}.
Since laser noise is the primary obstacle preventing current atomic clocks from operating at the SQL with long probe durations, dynamical decoupling has been proposed as a means of dealing with laser noise.

A well-established dynamical decoupling pulse sequence known as CPMG is a technique that is a quantum analog of the classical lock-in amplifier \cite{carr1954effects,meiboom1958modified}. To distinguish this from the new dynamical decoupling algorithm introduced below, in the following we refer to CPMG as narrowband dynamical decoupling (NBDD).  An ion in $\ket{\uparrow}$ is prepared in a superposition state $(\ket{\uparrow}+\ket{\downarrow})/2$ using a $\pi/2$ pulse. During the probe, a series of regularly spaced $\pi$ pulses are used to modulate the sign of the sensitivity to atomic or laser frequency fluctuations. A final $\pi/2$ pulse completes the pulse sequence. The frequency of the $\pi$ pulses is twice the frequency of the signal for which the sensitivity is improved (Fig.~\ref{fig:broadbandDD}b). This method was demonstrated on a $^{88}$Sr$^{+}$ ion by \citet{Kotler2011} and analyzed in the context of dark matter searches by \citet{aharony2021constraining}.

In the present work, we develop a new, broadband dynamical decoupling (BBDD) scheme that allows us to probe higher dark matter masses and resolves disadvantages of DS and NBDD schemes. We review each scheme. Then, we evaluate the performance of all three schemes in realistic numerical simulations based on the thorium nuclear isomer transition. The simulation results are compared to contemporary dark matter search experimental results and other proposals. Finally, an alternative nuclear excitation scheme in thorium is presented.

\section{Results}

\begin{figure}
\begin{center}
\includegraphics[width=1.0\columnwidth]{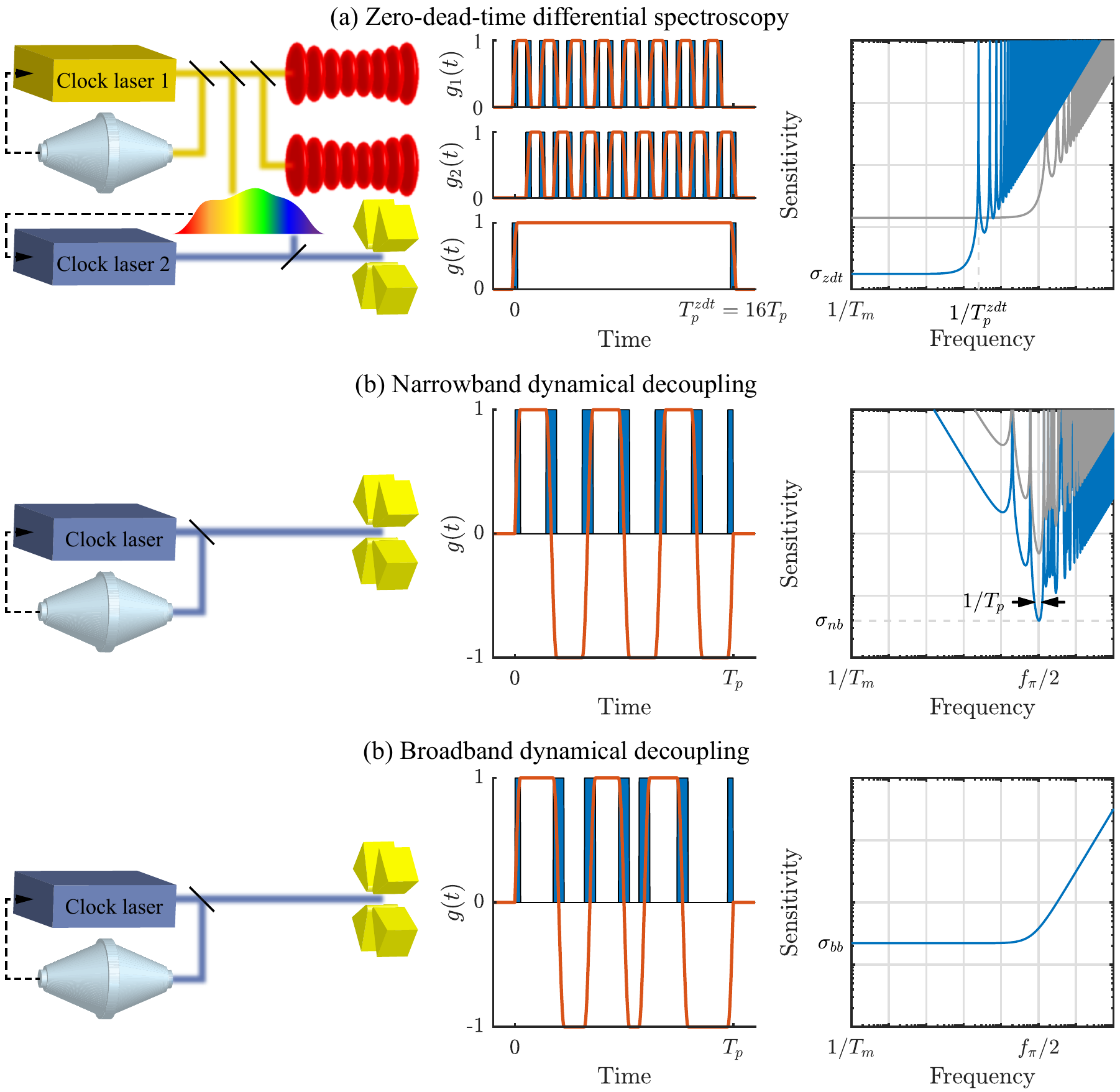}
\vspace{-0.7cm}
\caption{\label{fig:broadbandDD}
    Diagrams of experimental setups (left), pulse sequences (middle), and examples of theoretical sensitivity plots (right) for different quantum metrology algorithms:
	(a)
	zero-dead-time differential spectroscopy,
	(b)
	narrowband dynamical decoupling, and
	(c)
	broadband dynamical decoupling.
	Clock lasers are depicted as blue and yellow boxes; optical cavities used to stabilize the clock lasers are depicted as light grey cylinders with conical ends; feedback paths are indicated by black dashed lines; ensembles of atoms confined in optical lattices are depicted as stacks of red disks; and ions confined in ion traps are depicted as four yellow trap electrodes.
	The pulse sequences depict $\pi/2$ and $\pi$ pulses as blue rectangles with time increasing from left-to-right. The sensitivity $g(t)$ of the pulse sequence to clock transition or laser frequency fluctuations as a function of the time of the frequency fluctuations during the pulse sequence is shown as a red line.
	The sensitivity plots show the theoretical minimum detectable amplitude of an oscillating signal as a function of the oscillation frequency, such that higher sensitivity is indicated by the curve being lower on the plot.  $T_m$ is the total measurement duration and $T_p$ is the duration of a single repetition of the pulse sequence which is followed by a projective measurement for the blue curves.  The grey curves show how the sensitivity changes when $T_p$ is decreased.
}
\end{center}
\end{figure}


\noindent \textbf{Differential Spectroscopy.} The sensitivity of differential spectroscopy (Fig.~\ref{fig:broadbandDD}a) to deterministic oscillations of the ion clock transition frequency $\nu(t) = \nu_0 + \nu_S \cos(2 \pi f t + \theta)$ is identical to that of standard Ramsey spectroscopy, except that in differential spectroscopy (DS) the ion clock probe duration $T_p$ is not limited by the laser coherence time.  Extending the probe duration results in improved sensitivity to low frequency oscillations and reduced sensitivity to oscillations at frequencies larger than $1/T_p$.  In the oscillation frequency range $1/T_m \ll f \ll 1/T_p$ where $T_m$ is the total measurement duration, the small signal fractional frequency sensitivity to deterministic oscillations of unknown phase $\theta$ is given by
\begin{equation} 
\sigma_{Ramsey}(f) = \frac{X}{2 \pi \nu_0 \sqrt{T_p T_m}} \ .
\end{equation}
For $X = X_{det,95\%} \approx 3.95$, the oscillation amplitude $\left| \nu_S \right| < \sigma_{Ramsey}(f) \nu_0$ with 95\% confidence \cite{Centers2019}.

The phase of the atomic superposition state in the rotating frame accumulated during the $j$-th probe is
\begin{equation}\label{eq:sensitivityFunction}
\phi_j = \int_{t_j-T_p/2}^{t_j+T_p/2} 2 \pi g(t) \nu(t) dt \ ,
\end{equation}
where $t_j = j T_p$ is the start time and $T_p$ is now the duration of the pulse sequence for a single probe.  The sensitivity function \cite{Quessada2003} $g(t) = 2 \lim_{\delta\phi \rightarrow 0} \frac{\delta p}{\delta\phi}$ determines the change in the atomic transition probability $p$ caused by a detuning $\nu(t) = \nu_S \cos(2 \pi f t+ \theta) + \nu_{N}(t)$ between the ion transition frequency and the laser frequency, where $\nu_{N}(t)$ is the frequency noise, including the broadband laser frequency noise and quantum projection noise.

The sensitivity function is time-dependent and can be set to any value between -1 and 1 by applying laser pulses that change the orientation of the Bloch vector.  A standard Ramsey sequence, for example, consists of two $\pi/2$ pulses separated by the total sequence duration $T$.  We use the convention that the first $\pi/2$ pulse rotates the Bloch vector by $\pi/2$ radians about the $x$ axis, and the second $\pi/2$ pulse rotates the Bloch vector by $\pi/2$ radians about the $y$ axis.  (Here, we assume that the durations of the laser pulses are much shorter than the total sequence duration and can be neglected.)  The Ramsey algorithm has a sensitivity function $g(t) = +1$ for the duration of the sequence and therefore is optimal for the detection of a time-independent detuning.  The sign of the sensitivity function can be flipped by applying $\pi$ pulses on the $y$ axis during a Ramsey sequence.

\noindent \textbf{Narrowband Dynamical Decoupling.} Differential spectroscopy experiences a decrease in sensitivity at higher frequencies, especially at peaks of insensitivity that occur at integer multiples of $1/T_p$, which is a result of an integer number of dark matter oscillations fitting into the probe time. Narrowband dynamical decoupling in the style of Ref.~\cite{Kotler2011} introduces pulses within the probe time to achieve sensitivity for higher dark matter masses. It also suffers from the same problem (Fig.~\ref{fig:broadbandDD}b). A downside of this measurement scheme is it offers increased sensitivity for signals with a narrow frequency range, determined by the time interval between $\pi$ pulses. These problems can be overcome by randomly choosing the frequency of $\pi$ pulses for each Ramsey pulse sequence. This is the pulse sequence we have adopted here and in the following refer to as narrowband dynamical decoupling (NBDD).

\noindent \textbf{Broadband Dynamical Decoupling.} For dark matter searches over a broad range of particle masses, corresponding to a large range of detuning oscillation frequencies $f_{DM}$, we propose a broadband dynamical decoupling algorithm. In this algorithm, $\pi$ pulses about the $x$ axis are inserted into a Ramsey sequence at random times with a distribution designed to maximize the sensitivity to the oscillation frequency range of interest. Broadband dynamical decoupling offers all of the same advantages as narrowband dynamical decoupling with different random $\pi$ pulse frequencies in each probe when searching for a signal with unknown frequency, but further relaxes the requirements on laser coherence.

\noindent \textbf{Dark Matter Search Sensitivity with Thorium.} We perform numerical simulations to determine the sensitivity of various quantum metrology algorithms to clock transition frequency oscillations induced by dark matter. In these simulations we include realistic noise sources and the stochastic nature of scalar field dark matter.

For a deterministic frequency oscillation at frequency $f$, the phase accumulated during the $j$-th probe.
\begin{equation}\label{eq:phiSdet}
\begin{split}
\phi_{S,j}^{det} (f) & = 2 \pi \nu_S \int_{t_j-T_p/2}^{t_j+T_p/2} g(t) \cos (2 \pi f t + \theta) dt \\
& = 2 \pi \nu_S \left[ g_{I,j} \cos\theta - g_{Q,j} \sin\theta \right]
\ ,
\end{split}
\end{equation}
where
\begin{equation}
\begin{split}
& g_{I,j}(f) = \int_{t_j-T_p/2}^{t_j+T_p/2} g(t) \cos (2 \pi f t) dt \\
& g_{Q,j}(f) = \int_{t_j-T_p/2}^{t_j+T_p/2} g(t) \sin (2 \pi f t) dt
\end{split}
\end{equation}
\noindent are the in-phase and quadrature components of the sensitivity function of the $j$-th probe. Here, integration is performed over the duration of the $j$-th probe.  Ultralight DM induces oscillations that are stochastic in nature with a non-zero linewidth \cite{Centers2019}.  In this case, the phase accumulated due to the dark matter signal is given by
\begin{equation}
\phi_{S,j} = \int_{t_j-T_p/2}^{t_j+T_p/2} 2 \pi g(t) \nu_S(t) dt
\end{equation}
where $\nu_S(t)$ is the detuning caused by the dark matter field.

The measured phase for the $j$-th probe $\phi_{M,j} = \phi_{QPN,j} + \phi_{LN,j} + \phi_{S,j}$ where $\phi_{QPN,j}$ and $\phi_{LN,j}$ are the contributions due to quantum projection noise and laser noise, respectively.  Motivated by Eq.~(\ref{eq:phiSdet}), at the end of a measurement campaign consisting of $N_p = T_m/T_p$ probes, we compute the measured signal $\phi_M$ as a function of the analysis frequency $f$ by adding the phases of each probe with modulation-frequency-dependent signs, so that the signal contribution adds coherently:
\begin{equation} \label{Measured Signal}
\begin{split}
& (\phi_M)^2 = \frac{1}{2} \left( \sum_{j=0}^{N_p-1} \textrm{sgn}(g_{I,j}(f)) \phi_{M,j} (f_{DM}) \right)^2 \\
& + \frac{1}{2} \left( \sum_{j=0}^{N_p-1} \textrm{sgn}(g_{Q,j}(f)) \phi_{M,j} (f_{DM}) \right)^2 \ .
\end{split}
\end{equation}
This equation defines a coherent analysis of all probes together and is new in this work. The contributions from noise sources are independent of the dark matter signal contribution and thus add incoherently.

The fractional frequency uncertainty to coherent oscillations of the transition frequency at $f_{DM}$ can be written as
\begin{equation}
\sigma(f) = \frac{1}{\nu_0} \sqrt{\frac{\ave{(\phi_{N})^2}}{ \ave{(\phi_{S} / \nu_S)^2} } } \ .
\end{equation}

The phase accumulated during each probe due to laser frequency noise follows Eq.~(3). For differential spectroscopy, due to the feedforward phase correction from the lattice clock, the effective laser phase noise is actually the quantum projection noise of the lattice clock. It is taken to be Gaussian white noise with an Allan deviation set by the standard quantum limit for the lattice clock, which is taken to be a strontium lattice clock composed of 1000 atoms with a probe time of $1$~s. For the dynamical decoupling algorithms, the laser frequency noise is taken to follow a $1/f$ power spectral density as is typical for cavity-stabilized lasers. The quantum projection noise in the clock contributes $\phi_{QPN,j}=\pm 1$, which is added to the laser frequency noise contribution to give the total phase accumulated due to noise in the $j^{th}$ probe $\phi_{N,j}$. The phase accumulated in each probe due to the dark matter signal $\phi_{S,j}$ is calculated accounting for the nonzero dark matter line width due to decoherence. Further details pertaining to obtaining $\phi_{S,j}$ and $\phi_{N,j}$ are given in Methods. We account for non-zero dark matter line width by analyzing over a range of analysis frequencies $f$ around the dark matter Compton frequency $f_{DM}$. Then, the computed measurement result is fitted to the expected lineshape.

We run a full-scale simulation for DS, NBDD, and BBDD. For each dark matter Compton frequency $f_{DM}$, we analyze $1000$ frequency points $f$ in the vicinity. The measurement time used is $T_m=10^6~\text{s}$, and with a spectroscopy pulse sequence duration of $T_p=1$~s, this translates into $10^6$ simulated probes per analysis frequency. We use time steps of $dt=0.1/f_{DM}$ and $dt = 0.01/f_{DM}$ for dark matter line shape and laser frequency noise computations respectively. The probe time used for differential spectroscopy, narrowband dynamical decoupling, and broadband dynamical decoupling searches are $T_p=100$~s, $T_p=1$~s, and $T_p=0.25$~s, respectively. For dynamical decoupling sequences, a $\pi$ pulse frequency $f_{\pi}=20~\text{Hz}$ is used for the broadband case, and random frequencies between $2~\text{Hz}$ and $5~\text{Hz}$ are used for each probe for the narrowband case. For each measurement protocol, we simulate at least $100$ measurements for dynamical decoupling algorithms and $1000$ measurements for the differential spectroscopy case and fit to obtain the maximum amplitude of the theoretical dark matter line shape with $95\%$ confidence.

\begin{figure}
\begin{center}
\includegraphics[width=1.0\columnwidth]{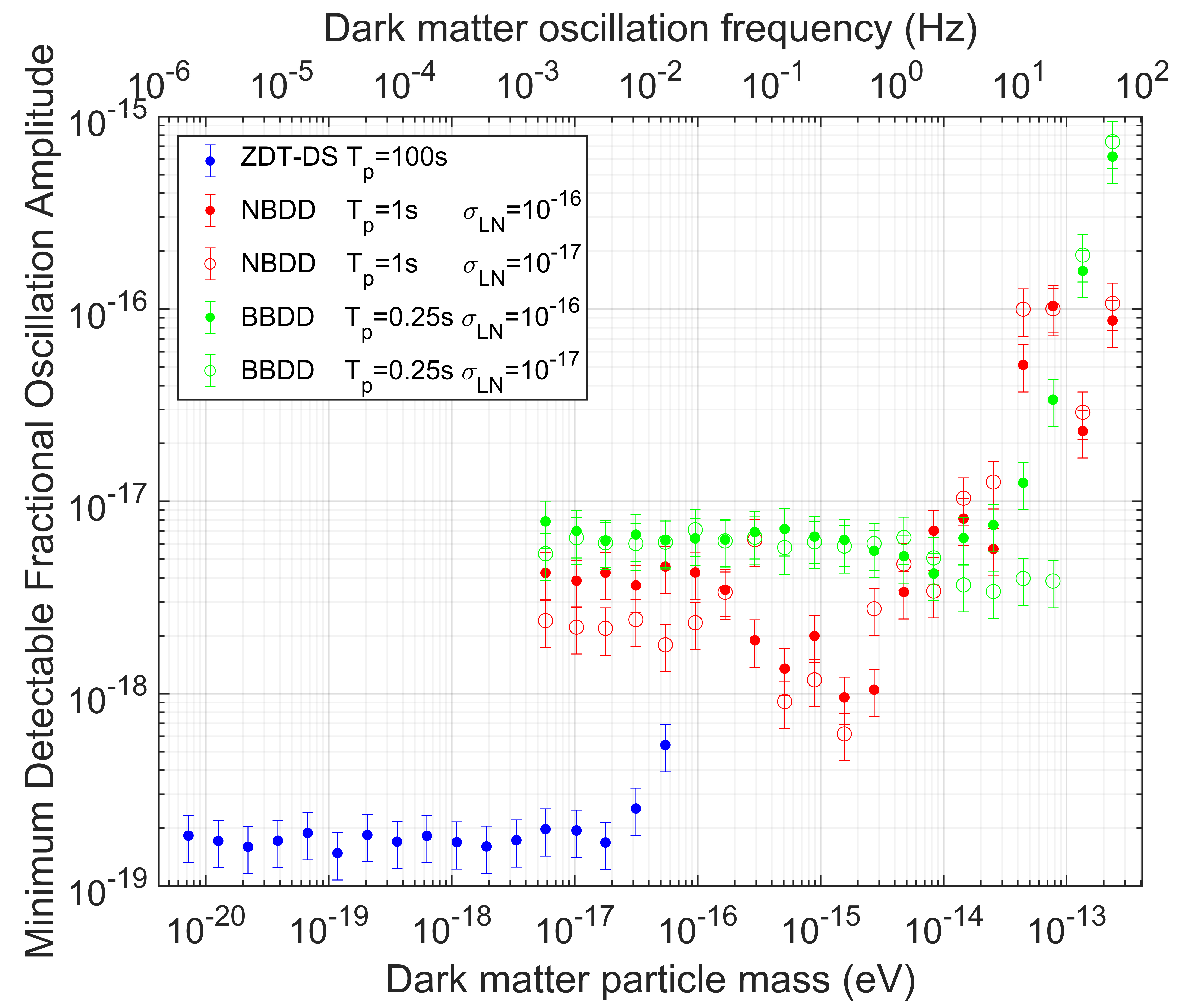}
\vspace{-0.7cm}
\caption{\label{fig:ExclusionPlot}
	Smallest detectable fractional frequency amplitude of oscillating clock frequencies due to dark matter for differential spectroscopy, narrowband dynamical decoupling, and broadband dynamical decoupling.  The sensitivity of the dynamical decouping algorithms depends strongly on the laser frequency noise, which is set to different levels for the filled and open circles.
}
\end{center}
\end{figure}

Fractional frequency uncertainty plots are presented in Fig.~\ref{fig:ExclusionPlot}. Differential spectroscopy rapidly loses sensitivity beyond $f_{DM}=1/T_p$ due to more than one dark matter oscillation fitting into the probe time. Narrowband dynamical decoupling achieves sensitivity for higher dark matter masses. Broadband dynamical decoupling offers comparable sensitivity with a lower probe time $T_p$ which is uniform throughout the large range of masses unlike either DS or NBDD. Most importantly, we clearly demonstrate that BBDD reaches 10$^{-13}$ DM mass range without a significant loss of sensitivity, critical to a recent proposal to search for DM halo bound to the Sun \cite{YuDai22}.

\begin{figure}
\begin{center}
\includegraphics[width=1.0\columnwidth]{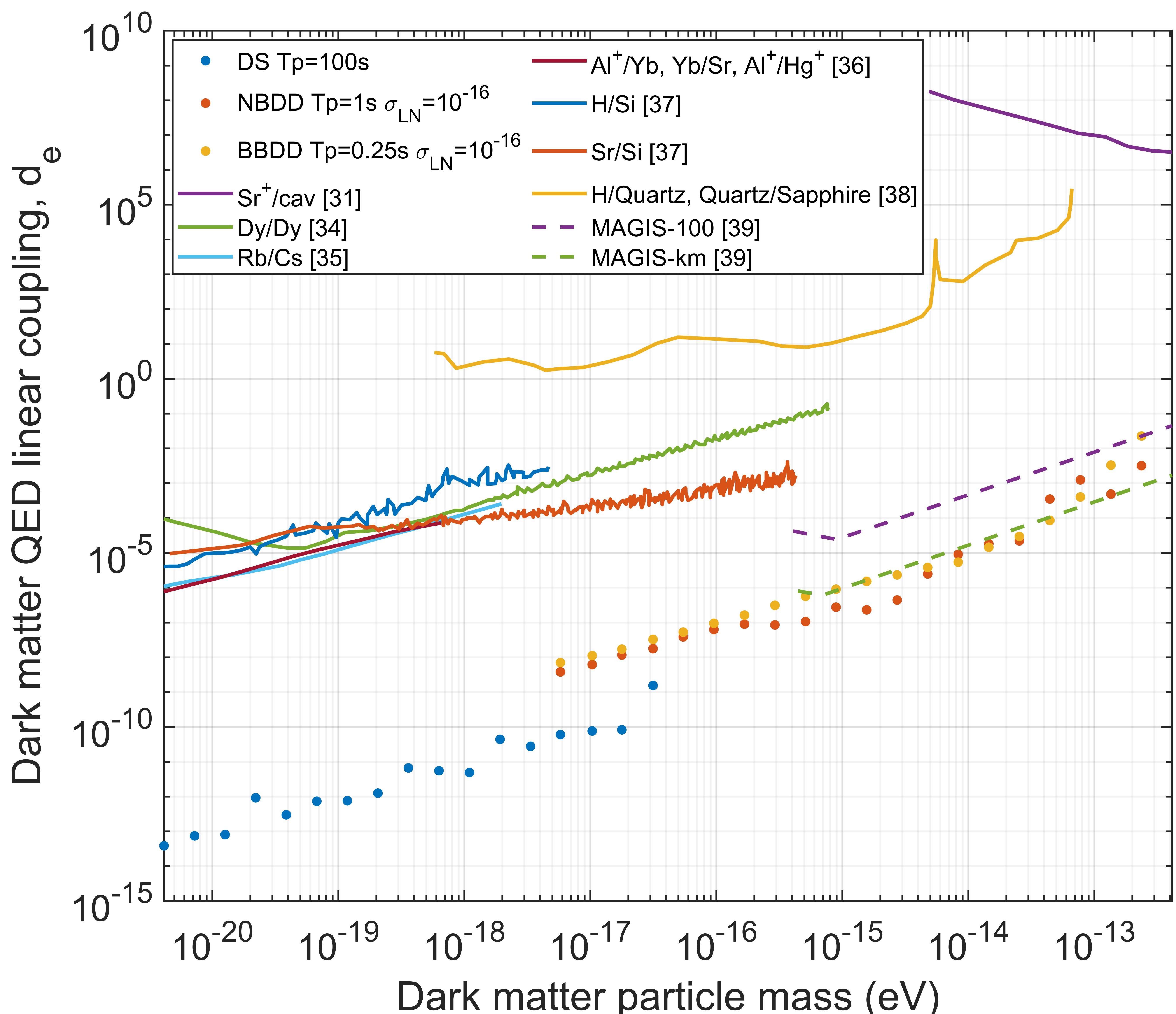}
\vspace{-0.7cm}
\caption{\label{fig:Fig4}
	Exclusion plot for the linear coupling constant of scalar dark matter to the QED electromagnetic field tensor.  Coupling constants larger than the solid lines have already been ruled out by previous experiments. Plotted circles show the simulated sensitivities of searches based on thorium nuclear clocks. The three quantum metrology algorithms considered here are shown. Other proposed experiments are shown as dashed lines \cite{van2015search,hees2016searching,boulder2021frequency,kennedy2020precision,aharony2021constraining,campbell2021searching,abe2021matter}. More details of these lines can be found in Ref.~\cite{antypas2022new}.
}
\end{center}
\end{figure}

An exclusion plot for $d_e$ presented in Fig.~\ref{fig:Fig4} shows an improvement in sensitivity by many orders of magnitude with nuclear clock experiments. For higher dark matter masses, only the proposed 1-km atom MAGIS interferometer \cite{antypas2022new} offers sensitivity similar to that of a table-top nuclear clock with dynamical decoupling.


\hfill

\hfill

\begin{figure}
\begin{center}
\includegraphics[width=1.0\columnwidth]{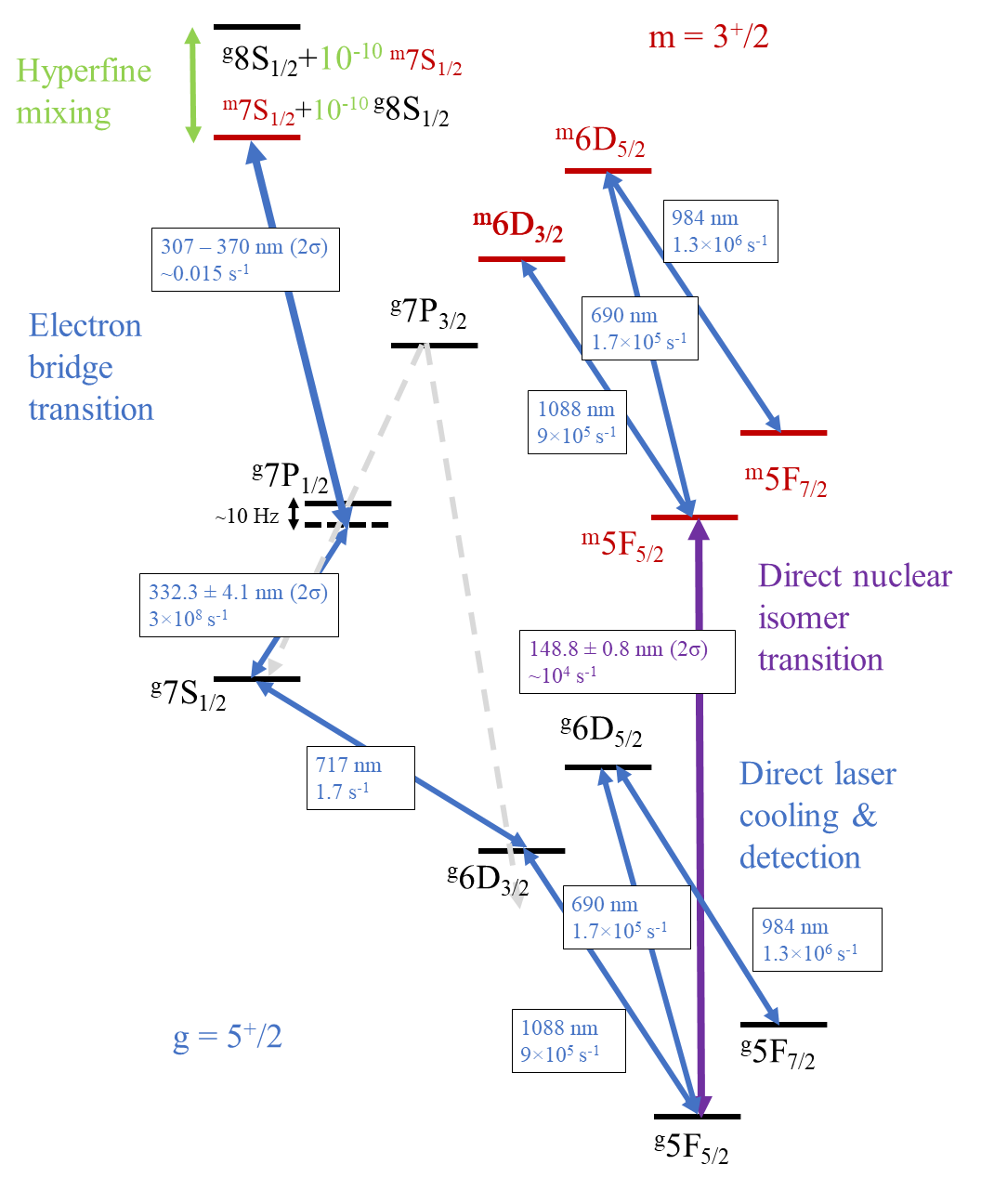}
\vspace{-0.7cm}
\caption{\label{fig:Fig5}
	(Color online)
    Energy level diagram of \ThIV isomer showing two methods of driving the clock transition.
}
\end{center}
\end{figure}

\noindent \textbf{Electron bridge two-photon spectroscopy.}  The thorium nuclear isomer transition has recently been measured to have a wavelength of $148.71(42)$~nm and an excited state lifetime of $670(102)$~s when embedded in a MgF$_2$ crystal \cite{Kraemer2022energy}.  Isolated \ThIV~ions are expected to have a longer lifetime.  The electronic and nuclear energy level structure of \ThIV~is shown in Fig.~\ref{fig:Fig5}.  For spectroscopy sequence durations approaching the lifetime limit of the isomer state, a natural choice of the electronic state for nuclear spectroscopy is the 5F$_{5/2}$ electronic ground state \cite{Campbell2012}.  In this case, the nuclear transition is most naturally driven using a single 149~nm photon.  Although significant progress has been made towards constructing narrow linewidth lasers at this wavelength \cite{Benko2014}, these lasers are very challenging and have limited up-time.

An alternative method is two-photon electron-bridge spectroscopy \cite{Porsev2010,Porsev2010a}.  Starting from the metastable $^g$7S$_{1/2}$ electronic excited state (where the $g$ superscript indicates the nuclear ground state), a two-photon transition to the nuclear excited state $^m$7S$_{1/2}$ can be driven using $^g$7P$_{1/2}$ as an intermediate state.  The second step of this transition relies on the fact that $^m$7S$_{1/2}$ contains a small admixture of the $^g$8S$_{1/2}$ state with an estimated amplitude of $\sim 10^{-5}$ \cite{Porsev2010a}.  The electron-bridge technique shifts the required laser wavelength from the very challenging VUV near 149~nm up to less challenging UV wavelengths of 269~nm and roughly 332~nm.

For precision nuclear spectroscopy using the electron-bridge technique, we propose detuning by $\sim 10$~GHz from the intermediate state and driving off-resonant two-photon transitions \cite{Wineland1998}.  For sufficiently low Rabi frequencies, the $^g$7P$_{1/2}$ is not populated and the spectroscopic coherence time will be limited by the 0.6~s electronic decay lifetime of the 7S$_{1/2}$ state \cite{Radnaev2012}.  This is sufficient for dynamical decoupling sequence durations up to a few hundred ms that are well-suited for searches for dark matter oscillations in the 0.1 to 1~kHz frequency range. Table~\ref{tab:COMBvsEB} compares various parameters in direct and electron bridge excitation methods.

\begin{table}
\caption{\label{tab:COMBvsEB}
Comparison of direct and electron bridge excitation methods.
}
\begin{ruledtabular}
\begin{tabular}{m{1.5in}
                >{\centering}m{0.8in}
                >{\centering}m{0.4in}
                >{\centering\arraybackslash}m{0.4in}}
					                & Direct    & \multicolumn{2}{c}{Electron Bridge} \\
\hline
Laser wavelength (nm)               & 149       & 269       & 332 \\
Laser power (W)                     & $10^{-6}$ & $10^{-8}$ & 1 \\
1/$e^2$ beam radius ($\mu$m)        & 10        & 30        & 3 \\
Detuning (GHz)                      &           & \multicolumn{2}{c}{10} \\
\hline
Min.~$\pi$ pulse duration (ms)   & 40        & \multicolumn{2}{c}{4} \\
Coherence time (s)                  & $10^{3}$ & \multicolumn{2}{c}{0.6} \\
Spont.~emis.~prob.~per~pulse &  & \multicolumn{2}{c}{0.02} \\
\end{tabular}
\end{ruledtabular}
\end{table}

\section{Discussion}

Ultralight scalar dark matter is highly theoretically motivated and there has been an increased interest in searches of such dark matter candidates. Thorium-based nuclear clock experiments will offer better sensitivity to ultralight scalar dark matter than any other existing or proposed experiments by many orders of magnitude for a large range of dark matter masses. While differential spectroscopy offers an excellent scheme for such experiments for a lower dark matter mass range, beyond $f_{DM}=1/T_p$, sensitivity is greatly reduced owing to many dark matter oscillations being packed within the probe time of the experiment. Dynamical decoupling sequences offer a solution and offer increased sensitivity for higher dark matter masses within the ultralight dark matter mass range, matched only by the proposed MAGIS-km experiment. Broadband dynamical decoupling in particular offers all these advantages, with sensitivity not being dependent on the probe time, 
so lower probe times can be used when employing this scheme. We have developed full scale simulations of clock experiments searching for ultralight scalar dark matter via alpha variations, which include dark matter decoherence, laser noise and quantum projection noise. Using these simulations, we have shown that broadband dynamical decoupling is less sensitive to laser noise and hence for a large range of masses offers similar sensitivity to ultralight scalar dark matter for laser noise with a laser noise level of $10^{-16}$ as it does with a laser noise level of $10^{-17}$.

The broadband dynamical decoupling technique proposed here could also find applications in laser frequency noise measurements similar to the work of \citet{Bishof2013}.  The frequency stability of lasers is typically characterized by measuring the beat note of the laser under test with a more stable reference laser.
However, for state-of-the-art clock lasers a more stable reference laser is often unavailable.  In this case, the frequency stability of the clock laser can be characterized using the atoms of an optical clock as the more stable reference, and the broadband dynamical decoupling algorithm provides a technique for performing laser frequency noise characterization at noise frequencies higher than is possible with conventional spectroscopy.

\section{Methods}

\noindent \textbf{Dark Matter Decoherence.} We treat the dark matter field as stochastic, rather than deterministic, based on the approach of Ref.~\cite{derevianko2018detecting}. This approach accounts for dark matter decoherence and hence is more appropriate for higher frequency dark matter, which corresponds to smaller coherence times. For stochastic fields, the power spectral density is not merely a delta function at the Compton frequency, but is distributed around it. It is proportional to the lineshape
\begin{equation} \label{Lineshape}
\begin{split}
& F(\omega) = (2\pi)^{-1/2} \tau_c \eta^{-1} e^{-\eta ^ 2} e^{-(\omega-2\pi f_{DM})\tau_c} \\
& \times \sinh(\eta \sqrt(\eta^2+2(\omega-2\pi f_{DM})\tau_c)) \ .
\end{split}
\end{equation}
\noindent Here $\eta$ is the ratio of galactic velocity and $v_{vir}$ which is the virial velocity of the dark matter halo, which is taken to be 1, while $\tau_c = 10^{6}/f_{DM}$ is the coherence time, and $f_{DM}$ is the dark matter Compton frequency. The analysis frequency $f$ is taken to correspond to all $\omega=2\pi f$ for which $F(\omega)$ is nonzero.

The power spectral density can be obtained using the line profile according to
\begin{equation}
\braket{|\tilde{\phi}_p|^2} = \frac{\pi N_f}{dt} \Phi_0^2 F(\omega_p) \ .
\end{equation}
\noindent Here $N_f$ is the number of points taken in frequency space to have 1000 points on the dark matter line, $dt = 0.1 / f_{DM}$ is the sampling interval chosen to be appropriately small, and $\Phi_0$ is the effective field amplitude related to the dark matter energy density $\rho_{DM}$ as $\Phi_0 = \frac{\hbar}{m_{\phi} c}\sqrt{2\rho_{DM}}$, where $m_{\phi}$ is the dark matter mass associated with the Compton frequency ($\omega_{DM}=m_\phi c^2/\hbar$).

Random coefficients in the frequency domain are generated by
\begin{equation}
p(\tilde{\phi}) = \prod_{p=0}^{N_f/2} \frac{1}{(\beta_p^{-1}\pi\braket{|\tilde{\phi}_p|^2})^{\beta_p}} \exp{-\beta_p \frac{|\tilde{\phi}_p|^2}{\braket{|\tilde{\phi}_p|}^2}}
\end{equation}
Finally, to obtain the amplitude $\nu_{S,j}$ and phase $\theta_j$ of the oscillating signal in the time domain, the discrete inverse Fourier transform $\phi_j = N_f^{-1} \sum_{p=0}^{N_f-1} \exp{(i\frac{2\pi}{N_f}jp)}\tilde{\phi}_p$ is taken, resulting in
\begin{equation}
\phi_{S,j} = \int_{t_j-T_p/2}^{t_j+T_p/2} 2 \pi g(t) \nu_{S,j} \cos(2 \pi f t + \theta_j) dt \ .
\end{equation}

\noindent \textbf{Dark Matter Lineshape.} The expected lineshape of the dark matter signal is obtained by convolving Eq.~\ref{Lineshape} with the absolute value of the sinc function:
\begin{equation}
F_c = 4\pi \int F(f - f') ~\text{abs}(\text{sinc}((f'-f-f_{DM})~ T_m))  df' \ .
\end{equation}
\noindent Here, $f'$ is the independent variable for integration and taken to be a large number of values ranging from $-1000/T_m$ to $1000/T_m$ for numerical evaluation of the integral. We take the number of values to be $N_i=10^5$ if the integration time $T_m$ is greater than the decoherence time $\tau_c$, otherwise $N_i=10^6$.

If the interrogation time is significantly lower than the coherence time, $T_m<0.1 \tau_c$, we don't use the lineshape  (\ref{Lineshape}) and instead just use the sinc function
\begin{equation}
F_c = \text{abs}(\text{sinc}((f-f_{DM})~ T_m))
\end{equation}
\noindent where $\omega_{DM}$ is the dark matter angular frequency.

The lineshape from the simulation $\phi_S$ comes from equation \ref{Measured Signal} where the contribution due to dark matter in the $j^{th}$ probe $\phi_{S,j}$ is given by Eq.~12. The simulated signal is fitted onto the expected line shape using the fmincon function in MATLAB, where the function minimized is the mean square error.
\begin{equation}
    \text{MSE}(p_1,p_2) = \textrm{mean}\left(\left(\sqrt{p_1^2~F_c^2 + p_2^2} - \phi_M\right)^2\right)
\end{equation}
\noindent The initial guesses for inputs $p_1$ and $p_2$ are taken to be $\text{max}(F_c)\text{max}(\phi_S)$ and $\sqrt{T_m/T_p} /4$ respectively. The fmincon function finds the values $p_1$ and $p_2$ which minimize the mean square error.

\noindent \textbf{Laser Noise.} Laser noise for dynamical decoupling sequences is produced with an approximate power law spectrum $S = S_0 \nu_{LN}^{\lambda}$ with $\lambda = -1$ for pink noise. The scale $S_0 = 1/\sqrt{2\log{2}}$ is chosen so that pink noise is produced with an Allan deviation of 1. A Mandelbrot state-space noise model is used. We generate a model of size $m=10$ to approximate the desired spectrom from a minimum frequency $f_{Min}=10^{-6}$~Hz to a maximum frequency $f_{Max}=10^3$~Hz. The elements of the transition matrix $A$, the input vector $B$, and the output vector $C$ are given by
\begin{equation}
A(i,i) = \exp \left(-\beta^{i-1} \frac{dt}{\alpha\tau}\right) \ ,
\end{equation}
\begin{equation}
B(i,1) = \frac{1-\exp(-\beta^{i-1}\frac{dt}{\alpha\tau})}{\beta^{i-1}} \ , \textrm{~and}
\end{equation}
\begin{equation}
\begin{split}
& C(1,i) = S_0 \frac{1}{\sqrt{2}} \alpha^{\frac{6-\lambda}{8}(f_{Min}dt)^{\lambda/2}} \\
& \times \prod_{j=1,j\ne i}^m(\alpha-1)\frac{\beta^{i-1}}{\alpha^m}\frac{\alpha\beta^{j-1}-\beta^{i-1}}{\beta^{j-1}-\beta^{i-1}} \ .
\end{split}
\end{equation}
\noindent Here, index $i$ goes from $1$ to $m$, $dt$ is the time step chosen to be an appropriately small fraction of the probe time, $\tau = 1/(2\pi f_{Min})$, $\beta=(10f_{Max}/f_{Min})^{1/m}$, and $\alpha=\beta^{-\lambda/2}$. The state vector $z$ for timestep $j$ is obtained from the one for timestep $j-1$ according to the update equation
\begin{equation}
z = A~z + B~r
\end{equation}
where $r$ is an $m$ element vector of normally distributed random numbers, which are regenerated for each $j$.  The state vector is used to generate the laser frequency at each timestep according to
\begin{equation}
f_{LN} = C~z \ .
\end{equation}
The laser noise frequencies are normalized as $\nu_{LN} = f_0 \times \sigma_{LN}\times f_{LN}$, where $f_0 = c / \lambda_0$ is the laser frequency, with $\lambda_0 = 148.8$~nm, and $\sigma_{LN} = 10 ^ {-16}$ or $\sigma_{LN} = 10 ^ {-17}$ is the Allan deviation of the flicker laser noise.

\section{Acknowledgements}

The authors thank J. Valencia and D. A. Hite for critical reading of the manuscript.  This work was supported by the Army Research Office (Grant Number W911NF-20-1-0135), the National Institute of Standards and Technology, the National Science Foundation Q-SEnSE Quantum Leap Challenge Institute (Grant Number OMA-2016244), the Office of Naval Research (Grant Number N00014-20-1-2513), and the European Research Council (ERC) under the European Union's Horizon 2020 research and innovation program (Grant Number 856415).
This research was supported in part through the use of University of
Delaware HPC Caviness and DARWIN computing systems: DARWIN - A Resource for Computational and Data-intensive Research at the University of Delaware
and in the Delaware Region, Rudolf Eigenmann, Benjamin E. Bagozzi, Arthi Jayaraman, William Totten, and Cathy H. Wu, University of Delaware, 2021, URL:
https://udspace.udel.edu/handle/19716/29071.

\section{Author contributions}

D.R.L., D.B.H., and M.S.S.~conceived the broadband dynamical decoupling algorithm.  D.R.L.~developed the simulation code for single processor computers.  M.H.Z.~parallelized the code and performed the simulations.  N.J.M.~developed and tested the Mandelbrot state-space noise model.  D.R.L. conceived the off-resonant two-photon excitation scheme.  All authors discussed the results and implications and contributed to writing and editing the paper.

\section{Competing interests}
The authors declare no competing interests.

%

\end{document}